\documentclass[aps,prl,preprint,superscriptaddress]{revtex4-2}

\usepackage{xcolor}
\usepackage{siunitx}
\usepackage{graphicx}
\usepackage{hyperref}
\hypersetup{
    colorlinks=true,
    linkcolor=magenta,
    filecolor=blue, 
    urlcolor=blue,
    citecolor=blue,
    pdftitle={Overleaf Example},
    pdfpagemode=FullScreen,
    }

\begin{document}


\title{Quantum Coherence of a Long-Lifetime Exciton-Polariton Condensate}


\author{Yannik Brune}
\affiliation{Department of Physics, Technische Universität Dortmund, Dortmund 44227, Germany}
\author{Elena Rozas}
\affiliation{Department of Physics, Technische Universität Dortmund, Dortmund 44227, Germany}
\author{Ken West}
\affiliation{Department of Electrical Engineering, Princeton University, Princeton, New Jersey 08544, USA}
\author{Kirk Baldwin}
\affiliation{Department of Electrical Engineering, Princeton University, Princeton, New Jersey 08544, USA}
\author{Loren N. Pfeiffer}
\affiliation{Department of Electrical Engineering, Princeton University, Princeton, New Jersey 08544, USA}
\author{Jonathan Beaumariage}
\affiliation{Department of Physics \& Astronomy, University of Pittsburgh, Pittsburgh, Pennsylvania 15260, USA}
\author{Hassan Alnatah}
\affiliation{Department of Physics \& Astronomy, University of Pittsburgh, Pittsburgh, Pennsylvania 15260, USA}
\author{David W. Snoke}
\affiliation{Department of Physics \& Astronomy, University of Pittsburgh, Pittsburgh, Pennsylvania 15260, USA}
\author{Marc Aßmann}
\affiliation{Department of Physics, Technische Universität Dortmund, Dortmund 44227, Germany}



\begin{abstract}
In recent years, quantum information science has made significant progress, leading to a multitude of quantum protocols for the most diverse applications. States carrying resources such as quantum coherence are a key component for these protocols. In this study, we optimize the quantum coherence of a nonresonantly excited exciton-polariton condensate of long living polaritons by minimizing the condensate's interaction with the surrounding reservoir of excitons and free carriers. By combining experimental phase space data with a displaced thermal state model, we observe how quantum coherence builds up as the system is driven above the condensation threshold. Our findings demonstrate that a spatial separation between the condensate and the reservoir enhances the state's maximum quantum coherence directly beyond the threshold. These insights pave the way for integrating polariton systems into hybrid quantum devices and advancing applications in quantum technologies. 
\end{abstract}


\maketitle

\section{Introduction}
The outstanding developments in quantum computation and quantum information science in a wide and diverse range of fields during the last few years led to a remarkable number of solutions for many arising challenges in fields such as quantum security \cite{Renner2008}, quantum machine learning \cite{Biamonte2017} and quantum optimization \cite{Moll2018}. The number of potential qubit platforms has significantly expanded, now encompassing systems as trapped ions \cite{Cirac1995,Ringbauer2022,Bruzewicz2019}, superconducting circuits \cite{Nakamura1999,Kjaergaard2020} or photonic systems \cite{Knill2001,Slussarenko2019}. Additionally, the losses and transmission rates of quantum links have been optimized substantially, allowing for longer distances to be covered \cite{Neumann2022,Hu2023,Chen2021,Lopez2022}. Despite these advancements, no single experimental platform proves ideal for all aspects of quantum technologies in quantum information science. Consequently, it is unavoidable to combine different 
platforms to achieve optimal performance in quantum technologies. Hybrid quantum devices that, e.g., can convert quantum information from a matter-based system to a photonic system, and vice versa, are therefore a key component for realizing such optimized approaches.
One hybrid system able to facilitate this conversion are exciton-polaritons, from now on referred to as polaritons, which arise from the strong coupling between excitons and confined cavity photons. Furthermore, due to their bosonic nature, these quasi particles show a wide range of nonlinear effects such as polariton lasing \cite{Laussy2004}, superfluidity \cite{Amo2009} and Bose-Einstein like condensation \cite{Kasprzak2006}.
Polariton condensates may carry quantum superpositions in the Fock basis, which can either be directly exploited as a resource or transferred into quantum entanglement or other non classical correlations \cite{Killoran2016,Ma2016}. This opens the path to a multitude of applications in quantum information science including quantum thermodynamics, quantum algorithms, quantum metrology or quantum phase transitions \cite{Streltsov2017}. Therefore, optimizing the amount of quantum superpositions in polariton condensates proves essential for applications in hybrid quantum technologies \cite{Ghosh2020,Opala2023,Ballarini2020,Boulier2020,Kavokin2022}.

Here, we create a polariton condensate by optically trapping non-resonantly excited, long-living polaritons within an annular-shaped laser beam to optimize the amount of quantum superpositions. We spatially separate the polariton condensate from the surrounding incoherent carrier reservoir, minimizing the interaction between them. In terms of $g^{(1)}$, it has already been demonstrated that this method improves the condensate's spatial \cite{Schmutzler2014} and temporal \cite{Askitopoulos2013,Orfanakis2021} coherence. To quantify the amount of quantum superpositions, we use the quantum coherence $C$, given by the squared Hilbert-Schmidt norm of the distance between the state's density matrix $\hat{\rho}$ and its closest incoherent counterpart $\hat{\rho}_{inc}$ \cite{Luders2021,Streltsov2017}:
\begin{equation}
    C(\hat{\rho})=||\hat{\rho}-\hat{\rho}_\text{inc}||_{HS}^2=\sum_{m,n,m\neq n}|\rho_{m,n}|^2.
\end{equation}
Quantum coherence consequently takes only the non-diagonal elements of the state's density matrix into account, making it complementary to the well-known photon correlation functions. Assuming the condensate can be described by a displaced thermal state, a superposition of both, coherent and thermal contributions, we demonstrate that the spatial separation between the condensate and the reservoir enhances the achievable quantum coherence of the system.

\section{Results and Discussion}
    \subsection{Characterisation of the condensation process}
      
      \begin{figure}[t]
            \centering
            \includegraphics[width=0.6\columnwidth]{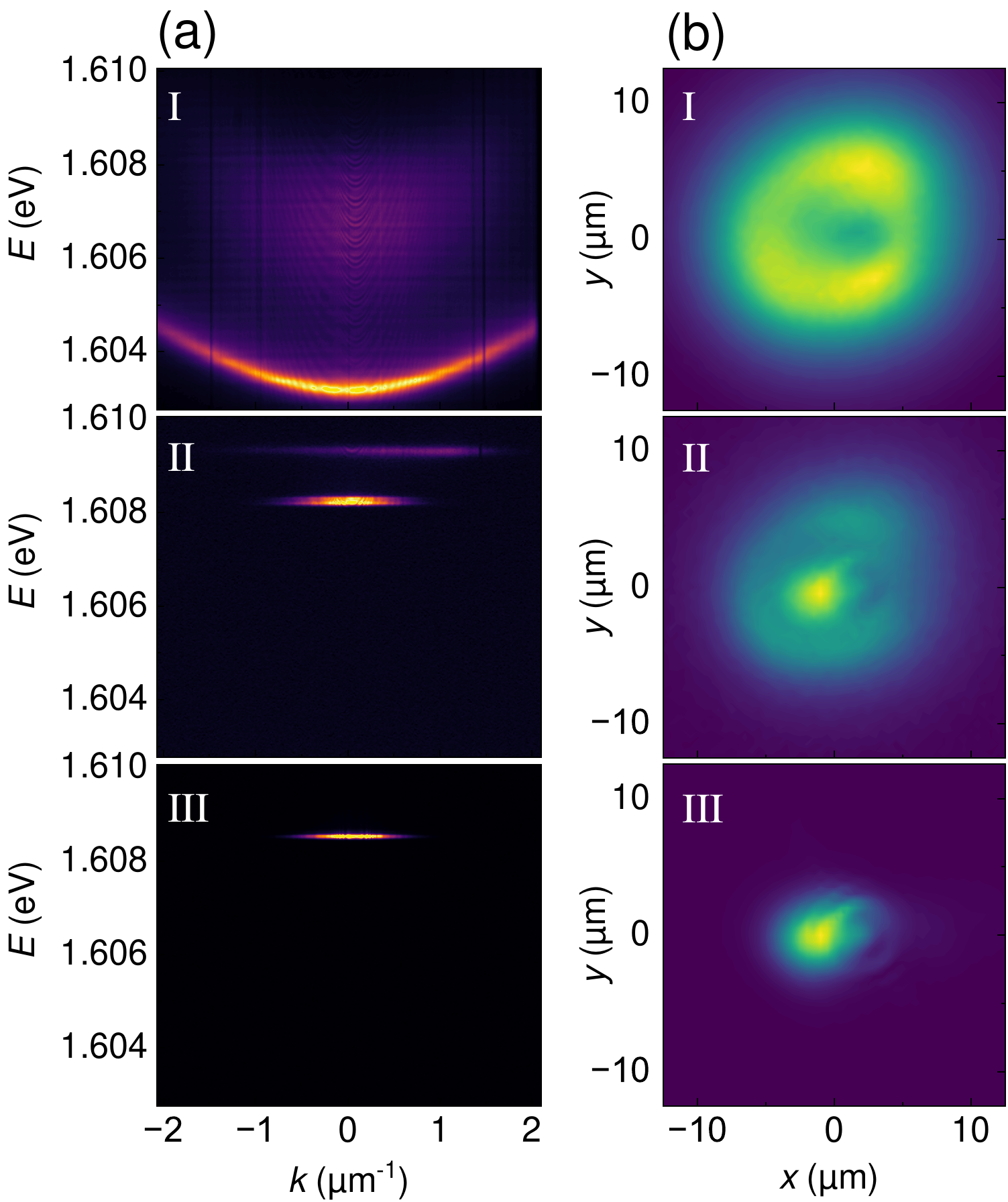}
            \caption{\textbf{Spectroscopic analysis of the power dependent polariton emission.}
            Polariton PL in momentum (a) and real space (b), at three different pump powers: $0.57\,P_\text{th}$ (I), $1.00\,P_\text{th}$ (II), and $1.35\,P_\text{th}$ (III). The images for case II are a superposition of consecutive images acquired under identical external conditions. In that case, the system shows a critical mode competition over time, switching between polariton emission within the circular barrier and a strong spontaneous condensate emission.}
            \label{fig:KandRspace}
        \end{figure}
        
    To ensure that the polariton condensate forms separately from the reservoir of carriers consisting of electrons, holes and also excitons, we create an optical trap using a spatial light modulator (SLM) to generate an annular-shaped continuous-wave laser beam with a diameter of $\SI{9.6}{\micro\metre}$ when focused on the sample. This configuration allows us to minimize the decoherence effects arising from carrier-polariton interactions. Fig. \ref{fig:KandRspace} shows the momentum and real space distributions of the resulting polaritons. Three different excitation densities are considered: (I) below the condensation threshold, (II) at the condensation threshold and (III) above the threshold. Below the threshold, the real space distribution displays polaritons remaining in the vicinity of the excitation laser, thus also exhibiting an annular-shaped emission pattern. Under these conditions, the ground state energy of the lower polariton branch (LPB) amounts to $\SI{1.603}{\electronvolt}$.
    As the pump power reaches the condensation threshold, the system becomes unstable and begins to switch between an emission either dominated by non-condensed polaritons or the condensate. Meanwhile, the momentum-resolved emission reveals a strongly blueshifted photoluminescence (PL) at $\SI{1.608}{\electronvolt}$, emerging from the condensate formed in the center of the optical trap. Although the polariton system is not yet stable, the first signatures of condensation start to appear. Therefore, we define this pump power as the power threshold for condensation (P$_{th}$). Once the system is above the onset for condensation (see panels III) the PL blueshift continues to increase with the excitation power. Throughout this process, the condensate remains at the center of the ring. Since its intensity shows a highly nonlinear increase, the emission from residual uncondensed polaritons vanishes in comparison.
    
          \begin{figure}[t]
            \centering
            \includegraphics[width=\columnwidth]{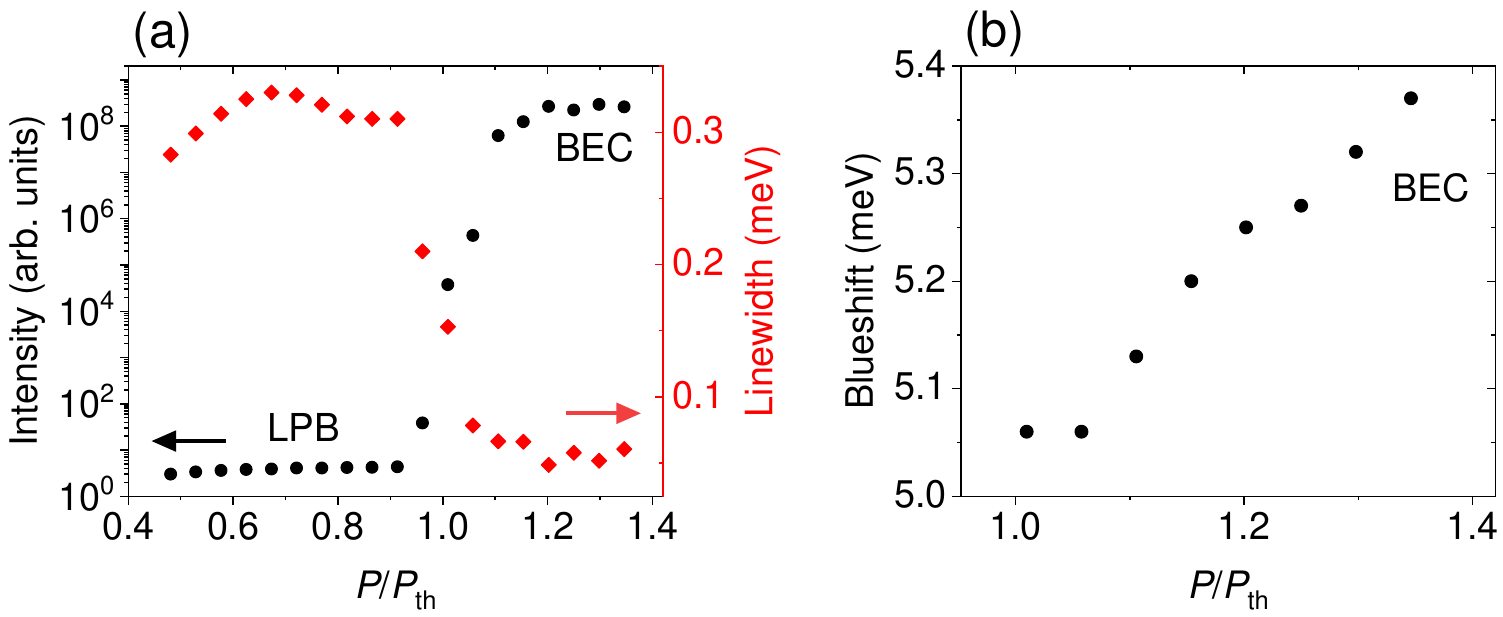}
            \caption{\textbf{Polariton condensation signatures.} (a) PL intensity and linewith as a function of P/P$_{th}$. (b) Corresponding blueshift extracted from the momentum-resolved emissions at $k=0$. LPB denotes spectra where the polariton system only occupies the ground state of the lower polariton branch. BEC denotes spectra where the system displays a polariton condensate state.}
            \label{fig:Nonlinear trend}
        \end{figure}
    
    To investigate the formation of the condensate in more detail, the intensity of the emission, i.e, the polariton occupation, its linewidth and blueshift are displayed in Fig. \ref{fig:Nonlinear trend} (a) and (b). All parameters are extracted by analyzing a region of interest in momentum space with a width of $\SI{1}{\per\micro\metre}$ and centered at $k=0$. As seen in panel (a), the intensity experiences a strong nonlinear increase by a factor of $10^8$ at the condensation threshold, while simultaneously the linewidth drops by a factor of six. Furthermore, when the system is excited at 1.35 P$_\text{th}$, the blueshift increases up to $\SI{5.4}{\milli\electronvolt}$.
    All analyzed parameters exhibit clear signatures of Bose-Einstein condensation, confirming our observations in Fig. \ref{fig:KandRspace}.

    It is worth noting that the blueshift experienced by the condensate is rather large compared to that reported in previous works on similar GaAs-based samples \cite{Nelsen2013,Sun2017,Alnatah2024}.
    This raises the question of the interactions at play and their respective contributions to the blueshift. In our case, the optical trap spatially separates the condensate from the reservoir of free carriers and bright excitons. However, as demonstrated in Ref. \cite{Myers2018}, long-lived dark excitons with k-vectors beyond the light cone, are able to move distances larger than $\SI{30}{\micro\metre}$. One must therefore assume that also a non-negligible number of dark excitons accumulates inside the center of the ring and interacts with the condensate \cite{Snoke2023}. Consequently, the system's quantum state must reveal the impact of these interactions on the  condensate's coherence \cite{Schmidt2019,Rozas2023}. 
    
    \subsection{Long timescale evolution of the polariton condensate subject to external noise}
  
\begin{figure}[t]
           \centering
           \includegraphics[width=0.8\columnwidth]{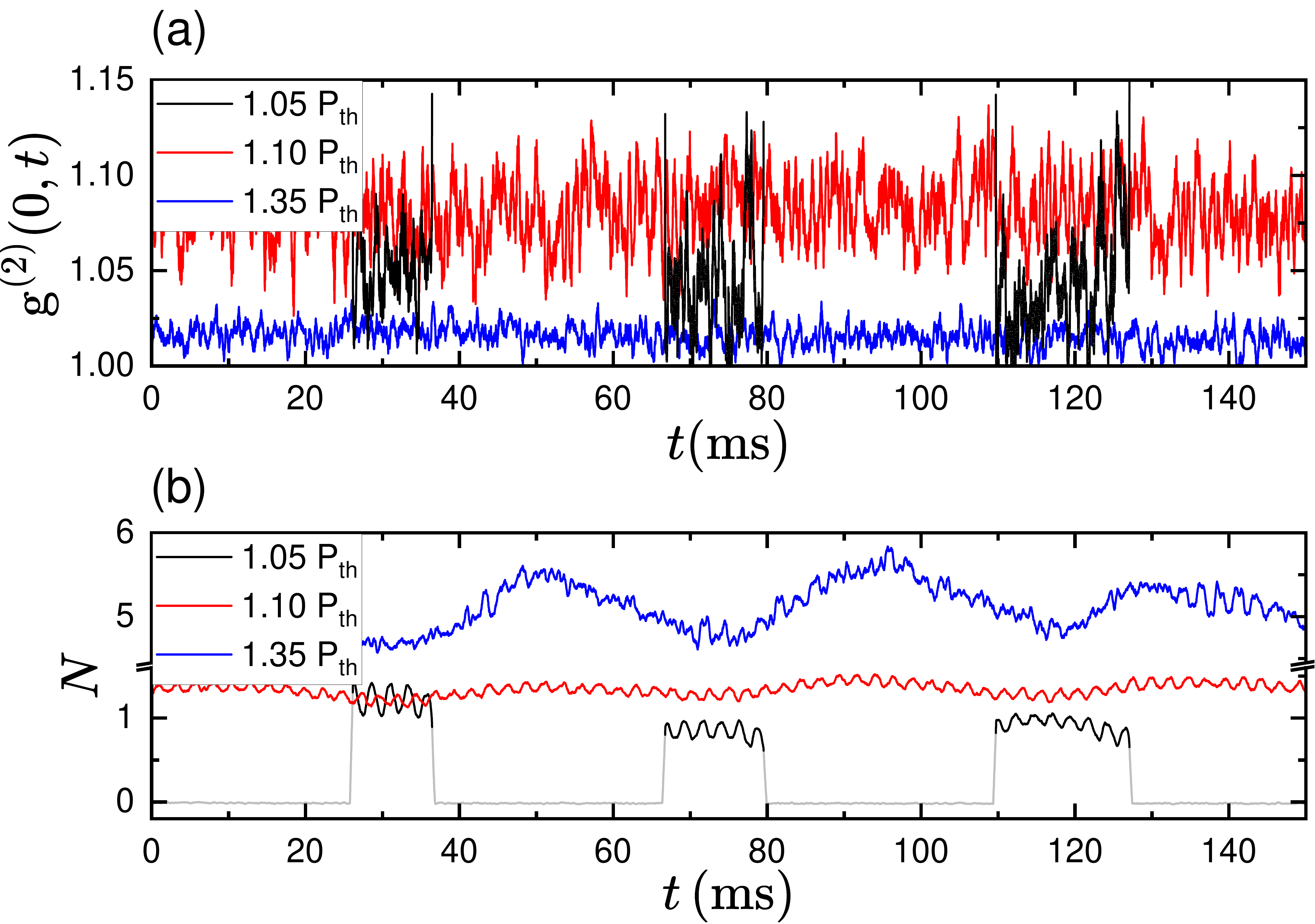}
           \caption{\textbf{Long scale time-resolved emission properties.} Long scale time resolved \textbf{(a)} second order correlation function $g^{(2)}(0)$ and \textbf{(b)} photon number of the polariton emission for the powers $1.05\,P_\text{th}$, $1.10\,P_\text{th}$ and $1.35\, P_\text{th}$. In the case of $1.05,P_\text{th}$ the system shows switching between an uncondensed state and a polariton condensate. For clarity, the uncondensed state photon number is greyed out and its $g^{(2)}(0)$-values are omitted.}
           \label{fig:N_g2_Timeresolved}
        \end{figure}

    In the last section we have demonstrated that the condensate is indeed separated from most of the carrier reservoir. However the influence of the interaction to dark excitons is not yet clear. To elucidate this point, we next examine the temporal evolution of the emission photon number and second order correlation function $g^{(2)}(0)$, calculated from single channel homodyne detection measurements. 
    The temporal dynamics of these two parameters are shown for the powers $1.05\,P_\text{th}$, $1.10\,P_\text{th}$ and $1.35\, P_\text{th}$ in Fig. \ref{fig:N_g2_Timeresolved}. At the lowest power of $1.05\,P_\text{th}$, immediately after the threshold, the system is still strongly affected by external noise. In consequence, the system switches repeatedly between an uncondensed and a condensed state. The transitions between both states are rapid, revealing the pronounced nonlinearities in the threshold region, which exhibit a high sensitivity to minor changes in the excitation. Whenever the system condenses, $g^{(2)}(0)$ drops immediately to around $1.04$, although it is still subject to significant low-frequency noise. During the time intervals where no condensate emission is detected, the measured signal is equivalent to vacuum. This is expected due to the absence of signal intensity in the optical mode of interest defined by the properties of the local oscillator. The local oscillator is on purpose set to the condensate mode and not overlapping with the uncondensed lower polariton branch. For clarity, the intervals where only vacuum is present are shown as greyed out for the photon number and omitted for $g^{(2)}(0)$.  
    
    As the excitation power increases to $1.10\,P_\text{th}$, the system stabilizes and stays continuously in a condensed state, exhibiting a slightly higher photon number in the condensed state compared to $1.05\,P_\text{th}$. Although we do not observe any more jumps in the photon number for this pump power, $g^{(2)}(0)$ does not decrease yet, probably due to the fact that polaritons still exhibit some thermal characteristics, as the power is not sufficiently far away from the threshold.
    Finally, at the maximal measured power of $1.35\,P_\text{th}$, the photon number rises to values between 4.5 and 6. At the same time, $g^{(2)}(0)$ decreases to a significantly lower value of $1.02$, with its uncertainty also decreasing significantly in magnitude. 
    
    The data for the powers of $1.10\,P_\text{th}$ and $1.35\,P_\text{th}$ exhibit, in addition to the already discussed trends, slow oscillations in time. These oscillations appear only in the photon number and not in $g^{(2)}(0)$. The reason for this behavior is an effective modulation of the signal outcoupling efficiency, caused by external mechanical noise. We evaluate $g^{(2)}(0)$ on timescales that are fast compared to the typical timescales of the external noise, so this extrinsic noise does not distort our measurements of  $g^{(2)}(0)$ \cite{Lueders2020}.

    \subsection{Estimation of the condensates quantum coherence}

\begin{widetext}
            \begin{figure}[t]
                \centering
                \includegraphics[width= \columnwidth]{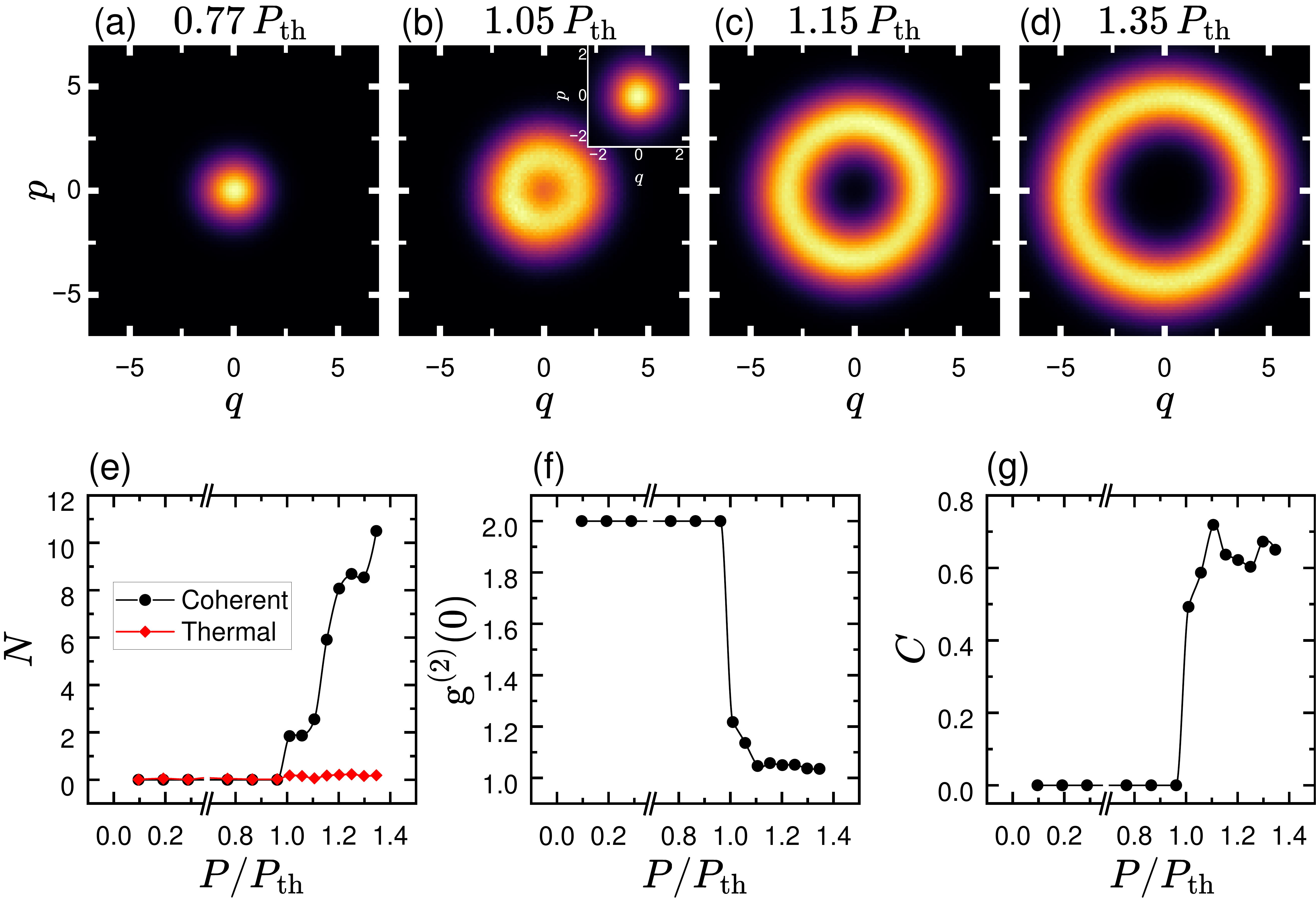}
                \caption{\textbf{Evolution of the polariton state in phase space.} (a)-(d) Phase averaged Husimi distributions for excitation powers of $0.77\,P_\text{th}$, $1.05\,P_\text{th}$, $1.15\,P_\text{th}$ and $1.35\,P_\text{th}$, displaying the transition from a vacuum state, characterized by a Gaussian probability distribution, to a displaced thermal state with a ring-like distribution. In case (b), the main figure only shows the condensate state distribution, the distribution of the whole recorded data set is given in the corresponding inset. (e) Coherent and thermal photon numbers extracted from the Husimi representations at different excitation powers. (f) Corresponding $g^{(2)}(0)$ as a function of the excitation power. (g) Calculated quantum coherence in dependence of the excitation power.}
                \label{fig:Husimi Results Grouped}
            \end{figure}
        \end{widetext}

    Finally, for a comprehensive understanding of the quantum state of our system, we discuss its coherence properties. By examining these properties, we gain deeper insights into the underlying mechanisms that govern the formation and stability of the polariton condensate. To achieve this, we measure orthogonal pairs of quadratures of the emitted light field using two-channel homodyne detection. The Husimi distributions then correspond to the 2D probability distributions of the recorded quadrature pairs (q,p). A more detailed description is given in the methods section. The resulting 2D distributions are exemplified in Fig. \ref{fig:Husimi Results Grouped} (a)-(d). Since the representation of the Husimi distribution is phase-averaged, this distribution is equivalent to the state's next incoherent counterpart $\hat{\rho}_\text{inc}$. It consists only of the state's main diagonal elements, while all off-diagonal elements vanish. For excitation powers around the threshold area, where the time-dependent emission shows contributions from condensed and uncondensed polaritons, as already highlighted in Fig. \ref{fig:N_g2_Timeresolved}, only the subset of data where a condensate is actually present is taken into account. This is especially the case for Fig. \ref{fig:Husimi Results Grouped} (b), which shows the subset of data where a condensate formed, while the inset shows the distribution of the full recorded data set.
   
    At $0.77\,P_\text{th}$, panel (a), the Husimi distribution exhibits a 2D Gaussian profile, indicating a vacuum state with no signs of condensation. When the system is driven above the threshold, the distribution evolves into a ring whose radius increases with power, indicating a system with increasing coherence. In this context, the entire polariton system can be seen as a superposition of both coherent condensed and thermal uncondensed polariton populations \cite{Adiyatullin2015,Deng2002,Vy2009}. The state of such a system can be described by a displaced thermal state, characterized by a coherent and thermal photon number.
    We can correlate the coherent photon number to the ring radius of the distribution, and its thermal photon number to the ring width. Both photon numbers are extracted from a fit to the Husimi Q distribution. Since the phase information is lost, it is sufficient for the fit to take only the distribution's radial dependency into account. To do so, we plot the probabilities gained from the Husimi distributions as a function of the radius. Then the fit is applied, allowing us to obtain the full information available in the 2D distribution.
    
    The resulting values are displayed in Fig. \ref{fig:Husimi Results Grouped}(e) as a function of the excitation power. Here it is important to remark that the coherent photon number immediately increases with power after the condensation threshold is exceeded. Meanwhile, the thermal photon number remains constant at a value close to zero. These results are attributed to the spatial separation between the condensate and polaritons at the excitation barrier, which is formed by thermal emission.
    Interpreting the state of the polariton condensate as a displaced thermal state, we estimate the coherence of our system by calculating both, $g^{(2)}(0)$ and its quantum coherence $C$ from the extracted photon numbers \cite{Luders2021,Klaas2018}. These estimations are presented in Fig. \ref{fig:Husimi Results Grouped}(f) and (g), respectively. Below the condensation threshold, our estimated $g^{(2)}(0)$ consistently exhibits a value of 2. We want to emphasize that our fitting method provides this value for a vacuum state and the uncondensed lower polariton branch does not overlap with the local oscillator. Nevertheless, previous works have exhaustively demonstrated that polaritons below the condensation threshold are generally characterized by a thermal state with $g^{(2)}(0)=2$ \cite{Adiyatullin2015,Deng2002,Vy2009}.
    
    As the condensation threshold is reached, $g^{(2)}(0)$ drops immediately to $1.2$, and with increasing power, further converges nonlinearly towards 1. 
    Conversely, as one focuses on the estimated quantum coherence displayed in panel (g), an abrupt jump from 0 to about 0.5 is observed. As power continues to rise, $C$ further increases, fluctuating between a value of 0.6 and 0.7. These fluctuations are mainly caused by minor variations in the thermal photon number, to which the quantum coherence is highly sensitive, unlike $g^{(2)}(0)$.
    
    Finally, we compare our estimated quantum coherence with previous measurements conducted by Lüders \textit{et. al.}, who used the same quantifier. In their case, shorter living polaritons in a lower Q-factor microcavity were excited by a Gaussian excitation spot of a diameter of $\SI{70}{\micro\metre}$ \cite{Luders2021}. In comparison, we achieve a quantum coherence three times higher than the previously observed maximum of $C=0.21$. 
    At first, this result may seem surprising. Although the polariton lifetime is much longer in our study which allows the condensate to thermalize more properly, also the condensate blueshift we observe is significantly larger compared to that earlier study. However, it should be noted that in our case the blueshift arises most likely from interactions between the condensate and a residual reservoir of long-lived dark excitons which also have sufficient time to thermalize.
    Therefore, it seems reasonable to assume that it is not the bare presence of a reservoir that reduces the coherence of a condensate, but rather its fluctuations.
    We conclude that increasing the polariton lifetime to bring the polariton condensate closer to a thermalized equilibrium state is significantly more beneficial to its quantum coherence properties than just separating it from incoherent reservoirs. It seems to be sufficient to separate the condensate from strongly fluctuating reservoirs as is the case for, e.g., reservoirs of free electrons and holes.  

\section{Conclusion}

In summary, our investigation reveals a significant enhancement in the amount of quantum coherence observed in polariton condensates compared to earlier studies. We ascribe this improvement to the long polariton lifetimes we are able to achieve, which in turn result in significantly better thermalization \cite{Alnatah2024}. Near the threshold, time-resolved measurements reveal an initially unstable polariton condensate, susceptible to external noise sources. Nonetheless, full stability is achieved already at $1.35\,P_\text{th}$, evidenced by $g^{(2)}(0)$ approaching unity. The extensive analysis of the system's coherence has demonstrated an effective increase of the coherent population resulting from first, the ability of long-living polaritons to overcome macroscopic distances and thermalize while maintaining their coherence, and second, minimizing the interactions with strongly fluctuating incoherent reservoirs. As a result, our measurements indicate a build up of quantum coherence significantly stronger compared to previous observations with Gaussian excitation profiles in moderate-lifetime samples \cite{Luders2021}. Taking these results into account, the influence of the residual excitonic reservoir could be further mitigated by increasing the excitation power and the size of the annular trap. However, most importantly our results show that the presence of incoherent reservoirs is not detrimental per se, which may open up the possibility to tailor them to even enhance the performance of polariton condensates in quantum technology tasks. Such insights hold promise for the integration of polariton condensates as low-threshold sources of resourceful quantum states for future hybrid quantum devices.

\section{Methods}
    \subsection{Sample}
        
         The sample used in this experiment is a high Q-factor $3\lambda/2$ optical microcavity containing 12 GaAs quantum wells (each $\SI{7}{\nano\metre}$ thick), embedded between two DBR mirrors. The QWs are arranged into three sets of four, with each set located at one of the three antinodes of the cavity. The DBRs consisting of 32(top) and 40(bottom) layers of AlAs/Al$_{0.2}$Ga$_{0.8}$As ensure a long polariton lifetime of $\sim$ 200 ps \cite{Steger2015}. The measurements have been performed in a region of the sample with a cavity-exciton detuning of $\delta \approx \SI{0}{\milli\electronvolt}$.

    \subsection{Experimental Setup}

        To create polaritons, the sample was cooled down to \SI{10}{\kelvin} in a cold-finger flow cryostat. We excited the sample nonresonantly at its first Bragg minimum, \SI{1.746}{\electronvolt}, with a continuous-wave laser manufactured by MSquared. The intensity distribution of the laser was reshaped into a ring using an SLM to modulate its phase front by imposing an axicon phase map. The diameter of the resulting ring-shaped optical trap was fixed at $\SI{9.6}{\micro\metre}$ during the experiments. A microscope objective with a numerical aperture of 0.26 was used for both, exciting the sample and collecting its photoluminescence. The PL was then filtered to allow only the polarization component parallel to the main axis of the cavity, which also corresponds to the polarization direction of the polariton condensate generated at high pump powers. 
        
\begin{figure}
            \centering
            \includegraphics[width=\columnwidth]{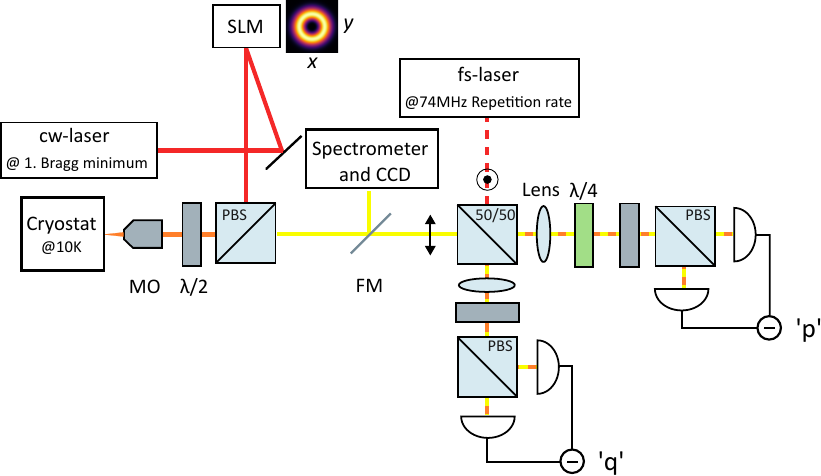}
            \caption{\textbf{Scheme of the experimental setup.}
            The sample is non resonantly excited in reflection geometry by a cw laser, spatially reshaped to an annular trap. The emission is then either analyzed using a combination of spectrometer and CCD or a two-channel homodyne detection setup. (SLM: spatial light modulator; PBS: polarized beam splitter; MO: microscope objective; FM: flip mirror; $\lambda/2$: half waveplate; $\lambda/4$: quarter waveplate)}
            \label{fig:SetupMain}
        \end{figure}

        For the analysis of the emission, two distinct techniques were used. First, the PL was focused on the entrance slit of a spectrometer and a charge-coupled device (CCD) to measure the spatial distribution and far-field angular distribution of the emission. Second, to investigate the quantum properties of the polariton emission, a two-channel homodyne detection setup was used to reconstruct the phase-space representation of the system's quantum state. To this end, the polariton PL was overlapped with a local oscillator, generated with a fs pulsed-laser beam, allowing a quadrature sampling rate of $\SI{74}{\mega\hertz}$, while maintaining a quadrature time resolution of around $\SI{1}{\pico\second}$. To ensure a stable orthogonality between the quadrature pairs (q,p) in both channels, the LO was delayed in one of the channels by adding an additional $\lambda/4$ plate. The full configuration of the setup is illustrated in Fig. \ref{fig:SetupMain}. For the detection, two high-speed balanced photoreceivers with integrated Si pin diodes and a bandwidth of $\SI{100}{\mega\hertz}$ were used. For a more detailed description of the used homodyne setup see Refs. \cite{Luders2018,Lueders2023,Luders2023CV}. 
       
        \subsection{Reconstruction and analysis of the quantum states}
        
        The calculation of the field quadratures, the Husimi-Q distribution states, as well as the followed analysis based on the model of a displaced thermal state, were calculated following the method described by Lüders \textit{et. al.} \cite{Luders2021}. Our measurements around the condensation threshold showed uncondensed and condensed polariton populations still coexisting in the system. Therefore, to proceed with the analysis, a minimum photon number of $0.6$ was selected to assume the system was indeed a polariton condensate. A further separation of the states is not considered since the calculated $g^{(2)}(0,t)$ shows no correlation with the observed temporal fluctuations in the photon number, as shown in Fig. \ref{fig:N_g2_Timeresolved}. 
        It is important to remark that the quantum coherence of the uncondensed polaritons could not be determined because I) the intensity of the emission was too low to be detected, therefore comparable to vacuum, and II) the strong blueshift of the emission prevents the energy of the LO from overlapping simultaneously with the condensed and uncondensed polaritons. Thus, only the condensate state was taken into account for the analysis of the system's usability in future hybrid quantum devices.

    \section{Data availability}
    Data underlying the results presented in this paper are not publicly available at this time. It may be obtained from the authors upon reasonable request.
    
    \section{Acknowledgments}
    This project was funded within the QuantERA II Programme that has received funding from the EU H2020 research and innovation programme under GA No. 101017733, and by the Deutsche Forschungsgemeinschaft (DFG) within the projects under GA No. 231447078 and 532767301. The Princeton University portion of this research is funded in part by the Gordon and Betty Moore Foundation’s EPiQS Initiative, Grant GBMF9615.01 to Loren Pfeiffer.
    
    \section{Author contributions}
    Y.B and E.R set up the setup, performed the experiments and analyzed the data. M.A. conceived and guided the research. K.W., K.B. and L.P. fabricated the sample which was designed by J.B., H.A. and D.S.
    The manuscript was written by Y.B, E.R and M.A incorporating feedback from all authors.
    
    \section{Competing interests}
    The authors declare no conflict of interests.

\providecommand{\noopsort}[1]{}\providecommand{\singleletter}[1]{#1}%

\end{document}